\newcommand{\beq}{\begin{equation}}
\newcommand{\eeq}{\end{equation}}
\documentstyle[preprint,aps]{revtex}
\tightenlines
\begin{document}
\draft
\title  { Magnetic-interference patterns in Josephson junctions 
 with $d+is$ symmetry}
\author{ N. Stefanakis$^{1,2}$, N. Flytzanis$^1$}
\address{ $^1$Department of Physics , University of Crete,
	P.O. Box 2208, GR-71003, Heraklion, Crete, Greece}
\address{$^2$Institute of Electronic Structure and Laser, Foundation for 
	Research and Technology - Hellas, GR-71110, P.O. Box 1527, 
	Heraklion, Crete, Greece}
\date{May 16, 1999}
\maketitle

\begin{abstract}

The magnetic interference pattern and the spontaneous flux in 
unconventional Josephson junctions of superconductors with $d+is$ 
symmetry are calculated for different reduced junction lengths and the 
relative factor of the $d$ and $s$ wave components. This is a time 
reversal broken symmetry state. 
We study the stability of the fractional vortex and antivortex which 
are spontaneously formed and examine their evolution as we change the 
length and the relative factor of $d$ and $s$ wave components. The 
asymmetry in the field modulated diffraction pattern exists for lengths as 
long as $L=10\lambda_J$.

\end{abstract}

\newpage

\section{Introduction}

In the past several years one of the main questions in the research
 activity on high-$T_c$ superconductors  has been the identification
of the order parameter symmetry \cite{scalapino,vanh,woll,tsuei,kirtley}. 
The most possible scenario is that the 
pairing state is an admixture of a dominant $d$-wave with some 
small $s$-wave component. This fact is a direct consequence of the
orthorhombic distortion of the systems which makes both the $d$-wave
and $s$-wave indistinguishable (they transform according to the
identity representation of the group).
There is a basic difference in the physics if one takes into account
 the phase difference between the two parts of the order parameter.
The mixing due to orthorhombicity predicts a $d+s$ or equivalently
$d-s$ order parameter. This has been analyzed within the  Ginzburg-Landau 
framework, valid close to T$_c$ \cite{betouras}.
Experimental observation of this possibility has been clearly realized
in photoemission experiments \cite{onellion} and the c-axis
 tunneling \cite{kouzn}.

In addition to the above work, calculations based in BCS weak-coupling 
theory \cite{musaelian,ren} predict that a mixed symmetry is 
realized in a certain 
range of interaction. This state has the time-reversal symmetry ${\cal T}$
broken. This symmetry is realized in bulk calculations only as
a consequence of the absence of any orthorhombic distortion (the Fermi  
surface is either circular or tetragonal in the particular examples) 
which favors a phase difference of $\pi/2$ between the
two components as opposed to $\pi$ in the presence of it.

The situation becomes more complicated if we consider surface effects. 
The observation of fractional vortices on the grain boundary in 
YBa$_2$Cu$_3$O$_7$ by Kirtley $et$ $al.$ \cite{kirtley2}, 
may indicate a possible violation 
of the time-reversal symmetry near grain boundary (because the boundary 
breaks the bulk orthorhombic symmetry).
Therefore it is interesting to study more this symmetry in the case
of interfaces. 

In the present paper we study the static properties of one dimensional 
junction which contains a twin boundary where the pair transfer integral
between the two superconductors has an extra relative phase in each twin. 
The maximum current $I_c$ that a junction can carry versus the external 
magnetic field $H$ in direction parallel to the plane of the junction 
is calculated by solving numerically the Sine-Gordon equation. The 
stability of fractional vortices $f_v$ or antivortices $f_{av}$ which are 
spontaneously formed as a consequence of the symmetry, is examined 
in the absence of current and magnetic field for different lengths and 
relative phases. 

In the ${\cal T}$-violated state the magnetic interference pattern 
as has been 
obtained by Zhu $et$ $al.$ \cite{zhu} in the short junction limit is 
assymetric. They conclude that for a long junction the magnetically 
modulated critical current is basically identical to the conventional 
0-0 junction due to the formation of the spontaneous vortex near the 
center of the junction. Our exact numerical calculations, show that 
there is a ``dip'' near the center of the diffraction patterns even for 
junctions as long as 10$\lambda_J$. 

The rest of the paper is organized as following. In section II we
discuss the Josephson effect for a mixed wave symmetry. In section III 
we present the results for the magnetic flux and the 
interference pattern. 
Finally, summary and discussions are presented in the last section.

\section{Josephson effect for a mixed wave symmetry} 

We discuss the Josephson coupling at the interface between two 
superconductors ($A$ and $B$) both with a two component order parameter 
$(n_1^{A(B)},n_2^{A(B)})$. We can think of the interface as a Josephson 
junction, so the Josephson current phase relation is \cite{bailey} 
\beq
J=\sum_{i,j=1}^2 J_{cij}\sin(\phi_i^B-\phi_j^A),
\eeq
where $J_{cij}$ is the coupling between the components $n_i^B$ on 
side $B$ and with $n_j^A$ on side $A$ [$n_j^{\mu} = 
|n_j^{\mu}|exp(i\phi_j^{\mu})$].
We consider some special cases.

($i$) For $d$-wave symmetry one component of the order parameter vanishes at 
the interface ($n_2=0$). The Josephson current density becomes 
$J=|J_{c11}|\sin(\phi+\pi)$ with $J_{c11}<0$.

($ii$) For $d+s$-wave we are restricted to the case where 
$\phi_1^A-\phi_2^A=\phi_1^B-\phi_2^B=\pi$ is fixed on both sides of the 
interface. The current density $J$ depends only on one phase difference 
through the interface, say $\phi=\phi_1^B-\phi_1^A$
\beq
J(\phi)=|\widetilde J_c|\sin(\phi+\theta) 
\eeq
\beq
\widetilde J_c=J_{c11}+J_{c22}-J_{c12}-J_{c21}
\eeq
with $\theta=0$ for $\widetilde J_c>0$ and $\theta=\pi$ for $\widetilde 
J_c<0$.

($iii$) For $d+is$-wave case the intrinsic phase difference within each 
superconductor $A$ and $B$ can be assumed to be
$\phi_1^A-\phi_2^A = \phi_1^B-\phi_2^B = \pi/2$.  
The current density $J$ is 
\beq
J(\phi)=\widetilde J_c \sin(\phi+ \theta)
\eeq
with 
\beq
\widetilde J_c=\sqrt{(J_{c11}+J_{c22})^2+(J_{c12}-J_{c21})^2},
\eeq
\beq
\tan(\theta)=\frac{J_{c21}-J_{c12}}{J_{c11}+J_{c22}}.
\eeq
We consider two superconducting sheets $A$ and $C$ which overlap 
for a distance $L$ with
the superconducting sheet $B$, in the $x$-direction, (as shown in Fig 1). All three 
superconducting sheets have dominant $d$-wave symmetry, with a small 
$s$-wave component. The angles between the crystalline $a$ axis of each 
superconductor $A,B,C$ with the junction interface are defined as 
$\theta_1, \theta_2, \theta_3$.
We describe the entire junction with width $w$ 
small compared to $\lambda_J$ in the $y$ direction, of length $L$ in the 
$x$ direction, in external magnetic field $H$ in the $y$ direction. 
The intrinsic phase difference  
$\theta(x)$ is $\phi_{c1}$ in $0<x<\frac{L}{2}$ and 
$\phi_{c2}$ in $\frac{L}{2}<x<L$. 
The superconducting phase difference $\phi$ across the junction is 
then the solution of the Sine-Gordon equation
\beq 
  \frac{d^2 { \phi}(x)}{dx^2} = \frac{1}{\lambda_J^2}\sin[{
\phi(x)+\theta(x)}] ,~~~\label{eq01} \eeq with the inline boundary condition
\beq \frac{d { \phi}}{dx}\left|_{x=0,L}\right. =\pm \frac{{
I}}{2}+H   ~~~~\label{eq02} \eeq
The Josephson penetration depth is given by 
\[
\lambda_J=\sqrt{\frac{\hbar c^2}{8\pi e d \widetilde J_c}} 
\] where $d$ is the sum of 
the penetration depths in two superconductors plus the thickness of the 
insulator layer. We also assume that $\widetilde J_c$ is constant within 
each segment of the interface.

To check the stability we consider small perturbations 
$u(x,t)=v(x)e^{st}$ on the static 
solution $\phi(x)$, and linearize the time-dependent Sine-Gordon 
equation to obtain:
\beq 
  \frac{d^2 v(x)}{dx^2}  +\cos[\phi(x)+\theta(x)] v(x)= \lambda v(x)
 ,~~~\label{eq10} \eeq 
under the boundary conditions $\frac{d v(x)}{dx}|_{x=0,L}=0$, where $\lambda=-s^2$. It is 
seen that if the eigenvalue equation has a negative eigenvalue the 
static solution $\phi(x)$ is unstable.

We can also compute the free energy of the solution for zero current
and external magnetic field 
\beq 
  F = \frac{\hbar \widetilde J_c w}{2 e} \int_{0}^{L} \left[ 1-\cos 
\left[ \phi(x)+\theta(x) \right]
+\frac{\lambda_J^2}{2} \left( \frac{\partial \phi}{\partial x} \right)^2 \right] dx
 ,~~~\label{eq11} \eeq 
Note that the no vortex solution $\phi=0$ everywhere is not a solution of 
this problem.

When $\phi_{c1}=\phi_{c2}=0$ we have the conventional $s$-wave junction.
In case $\phi_{c1}=0$, $\phi_{c2}=\pi$ we have the $d$-wave or $d+s$-wave 
junction. The above cases have time reversal symmetry 
(${\cal T}$-conservation).
When $\phi_{c1},\phi_{c2}$ are slightly different from $0$ and 
$\pi$, we have the $d+is$-wave pairing, which is a broken time reversal 
symmetry state (${\cal T}$-violation).
In this work, the particular parameters we use are $\phi_{c1}=0.01\pi$, 
$\phi_{c2}=1.08\pi$, and the pairing state is $d+is$.

\section{Spontaneous magnetic flux and interference patterns for the 
${\cal T}$-violating pairing state} 

In Fig. 2 we plot the maximum current for a symmetric $0-\pi$ junction
as a function of the magnetic flux $\Phi$ (in units of $\Phi_0=\frac{h 
c} {2 e}$) for different 
junction lengths: (a) $L=10$, (b) $L=4$, (c) $L=2$, (d) $L=1$ 
($\lambda_J=1$).
 The circles and squares in this figure correspond to the fractional vortex 
($f_v$) and antivortex ($f_{av}$) branch. For most of the range of 
existence of $f_v$ ($f_{av}$) the magnetic flux is positive (negative) 
while there is a small region where it turns into antivortex (vortex).
 Similar calculation has been done \cite{xu}, \cite{kirtleymoler} 
 who considered  
the $f_v$, since in this case the plot is symmetric in $H$. As we can 
see,  there is a ``dip'' at $\Phi=0$, for lengths as long as $L=10$. 
In Fig. 3 we present our calculations for the ${\cal T}$-violation case where 
$\phi_{c1}=0.01\pi$ and $\phi_{c2}=1.08\pi$.
We also plot for $L=1$ the analytical 
result (solid line) of Zhu $et$ $al.$ \cite{zhu}. In contrast to the pure 
$d$-wave case, for small lengths this pattern is assymetric and 
the ``dip'' in the 
maximum current does not occur at $\Phi=0$, but at a finite $\Phi$ value. This behavior also exists for 
lengths as long as $L=10$. 

If we plot $I_c$ vs 
$H$ (and not $\Phi$) then the two branches in Fig. 3, will be 
almost coincident and one might draw the conclusion that the behavior 
for a long junction is the same independent of the symmetry. The proper 
quantity to consider though is the total magnetic flux which includes 
both the contribution from the external field and the induced self 
field. It should be remarked that for an $s$-wave junction the relation 
between $\Phi$ and $H$ is linear for small $H$ so that the plot of 
$I_c$ vs $H$ or $\Phi$ does not show any differences for small $H$. 
For higher $H$ however the overlapping branches (for long $L$) are 
unfolded. In the case of a different symmetry even the small $H$ form 
can change due to the existence of spontaneous magnetization.
In this case, if we consider the zero current 
solutions, and vary the magnetic field, both $f_v$, $f_{av}$ are
stable, whereas 
for $L=4$, $L=2$, $L=1$ the stable regions in the magnetic field are 
separated by the unstable ones, and this behavior persists in both the $f_v$,
$f_{av}$ cases. Also in 
the short junction limit $(i.e.$ $L=1)$ these two branches coincide at the 
maximum current.

Figure 4 addresses the question of spontaneous flux generation in 
junctions with broken time reversal symmetry (${\cal T}$-violation) 
as a function of 
the reduced length ($L$) and the relative factor of $s$ and $d$ 
components.
The long dashed line is the result of \cite{zhu} which compares with 
 our numerical result (solid line). Both cases have $\phi_{c1}=0.01\pi, 
\phi_{c2}=1.08\pi$.  
We have also used two other values for $\phi_{c2}$ i.e. $0.9\pi$ (dotted 
line) and $0.8\pi$ (dashed line). 
We conclude that as we decrease the value of $\phi_{c2}$ the fractional vortex $f_v$ 
tends out to be a $2\pi$ vortex, whereas the fractional antivortex 
gradually loses its flux content. 

In Fig. 5a we have plotted the magnetic flux 
$\Phi$ (solid line, $\Phi_0=1$) versus the value $\phi_{c2}$ for $L=10$ and $H=0$. 
We see that as we decrease $\phi_{c2}$ the magnetic flux increases 
linearly for the $f_v$ branch from $\Phi \approx 0.45$, 
for $\phi_{c2}=1.08\pi$
to $\Phi \approx 1$ for $\phi_{c2}=0$. But this last point is 
unstable, as seen by the stability analysis from which the lowest 
eigenvalue is also displayed (light line). This is expected since it goes to a 
point in the unstable ($1,2$) branch of the usual $s$-type junction  
\cite{cap}.
The $f_{av}$ branch  increases linearly its flux as we decrease 
$\phi_{c2}$ and goes to the 
stable ($0,1$) branch. Here we follow the notation of \cite{owen}.
This linear dependence of $\Phi$ from $\phi_{c2}$ can also be 
seen in the analytical result of \cite{zhu} for large lengths, where the 
approximation they made is valid.
On the other hand as we increase $\phi_{c2}$ from $1.08\pi$ to $2\pi$, 
the $f_v$ branch decreases its flux and goes to the stable ($0,1$) branch, 
while the $f_{av}$ branch goes to the unstable ($-2,-1$) branch.
When we change $\phi_{c1}$ and keep $\phi_{c2}=0$, from $0$ to $2\pi$, the 
($0,1$) branch goes to $f_v$ and then to the unstable ($1,2$), while the 
unstable ($-2,-1$) branch goes to $f_{av}$ and then to the stable ($0,1$).
The situation is a little bit different for small lengths as can be seen 
from Fig. 5b where $L=1$, $H=0$. Here $f_{av}$ branch is unstable for 
$\phi_{c2}=1.08\pi$ and by decreasing $\phi_{c2}$ it gets stabilized, 
but $f_v$ branch is stable for $\phi_{c2}=1.08\pi$ and then it becomes 
unstable. We can see that from Fig. 6 where we plot the ratio $F/F_0$ of the 
free energy of the state with some spontaneous flux to the state with no 
flux. This ratio becomes larger than one as we decrease the $\phi_{c2}$, 
for the $f_v$ branch, for small lengths. On the other hand, when $F/F_0<1$ the 
no flux state is metastable and the final state will be the one with spontaneous 
flux. Notice that the magnetic flux remains 
almost constant - almost zero -  which can be expected since we are 
in the short junction limit where self currents are neglected.

\section{Conclusions} 
We have studied the static properties of a one dimensional junction 
with $d+is$ order parameter symmetry. The magnetic interference 
pattern is asymmetric, and there exist a ``dip'' near $\Phi=0$ for 
lengths as long as $10\lambda_J$. 
The diffraction pattern of a 
junction can give us information about the pairing symmetry, at least
where junctions are formed.

We have followed the evolution of spontaneously formed vortex and 
antivortex solutions for different mixing between the $s$ and $d$ 
components of the order parameter. We have shown that for small lengths 
the fractional vortex becomes unstable as we decrease the extra phase of 
the pair transfer integral in the right part of the junction. 
We conclude that when a mixing state symmetry is realized, the fractional 
vortex and antivortex solutions evolve differently and this characterizes the 
$d+is$-wave pairing. We expect these findings to hold even if a bulk
$d+s$ state evolves continuously as a function of distance from the 
interface to a $d+is$ one, 
as long as the there is a well defined  area close to the
interface where the time reversal symmetry is not conserved and the
junction is formed.

\begin{figure}
\caption{The one dimensional junction geometry. The dashed line marks the 
twin boundary.}
\label{fig1}
\end{figure}

\begin{figure}
\caption{Critical current $I_c$ versus the magnetic flux $\Phi$ (in 
units of $\Phi_0$)  
for a symmetric $0-\pi$ junction, for different junction lengths: (a)
$L=10$, (b)$L=4$, (c)$L=2$, (d)$L=1$.}
\label{fig2}
\end{figure}

\begin{figure}
\caption{Critical current $I_c$ versus the magnetic flux $\Phi$ 
for a junction with $d+is$ symmetry, $\phi_{c1}=0.01\pi$, $\phi_{c2}=1.08\pi$, for different 
junction lengths: (a) $L=10$, (b)$L=4$, (c)$L=2$, (d)$L=1$.}
\label{fig3}
\end{figure}

\begin{figure}
\caption{The spontaneous magnetic flux $\Phi$ as a function of the 
reduced junction length $L$, for different values of the intrinsic
phase  $\phi_{c2}$ in the right part $L/2<x<L$ of the junction and 
$\phi_{c1}=0.01 \pi$.}
\label{fig4}
\end{figure}

\begin{figure}
\caption{ The evolution of the fractional vortex $f_v$ and antivortex 
$f_{av}$ 
as a function of $\phi_{c2}$, for two different lengths (a) $L=10$, (b) 
$L=1$. The stability of theses branches is also denoted by the lowest 
eigenvalue $\lambda_1$ of the linearized eigenvalue problem. Note that 
the double arrow connects the flux with its stability curve.}
\label{fig5}
\end{figure}

\begin{figure}
\caption{The ratio of the free energy $F/F_0$ as a function of the 
reduced junction length $L$, for different values of $\phi_{c2}$: (a) 
$f_v$, (b) $f_{av}$.}
\label{fig6}
\end{figure}

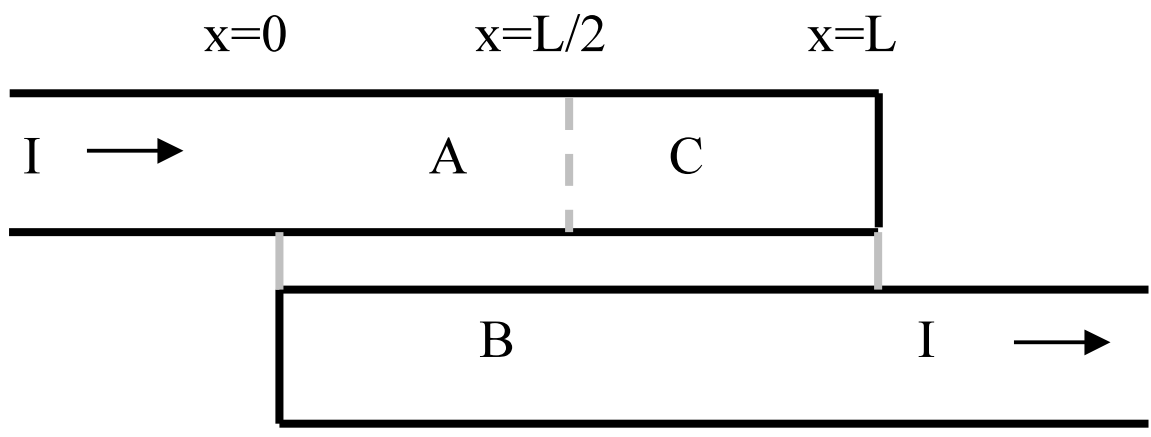
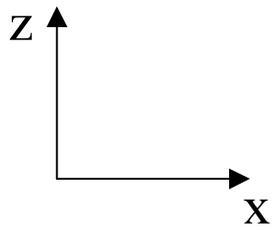
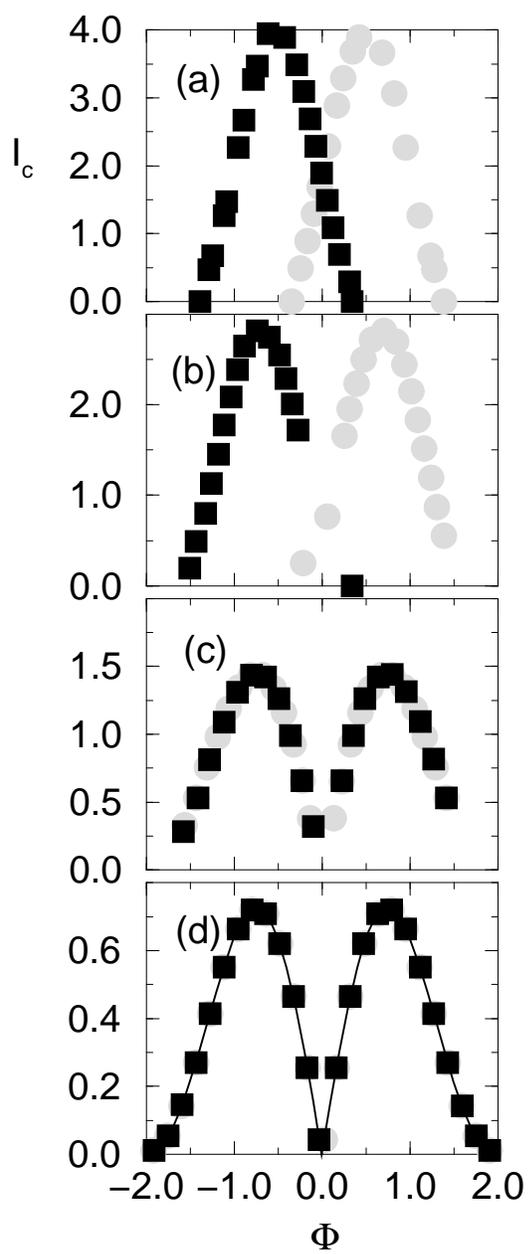
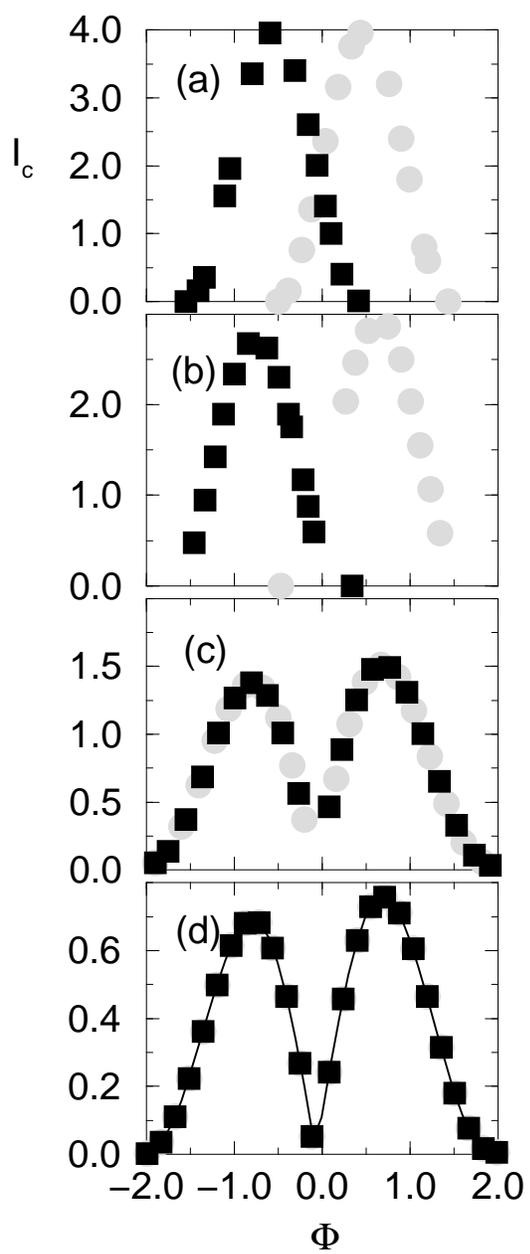
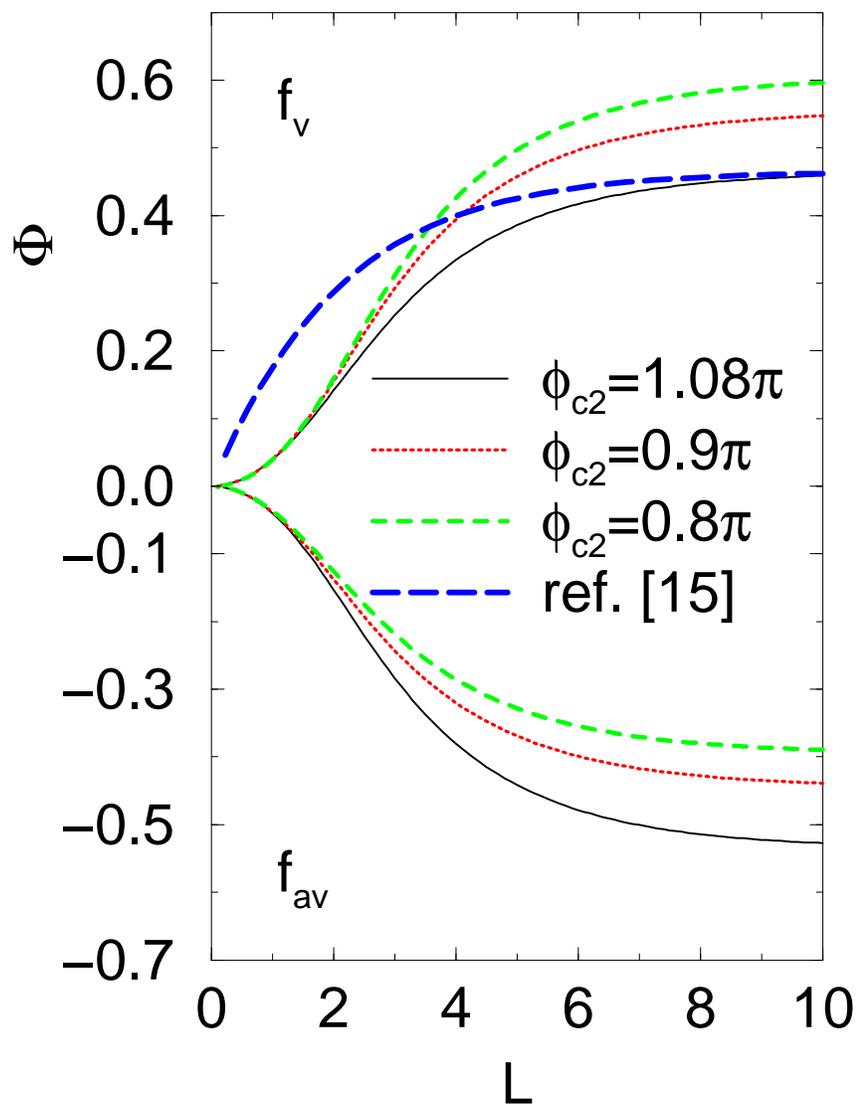
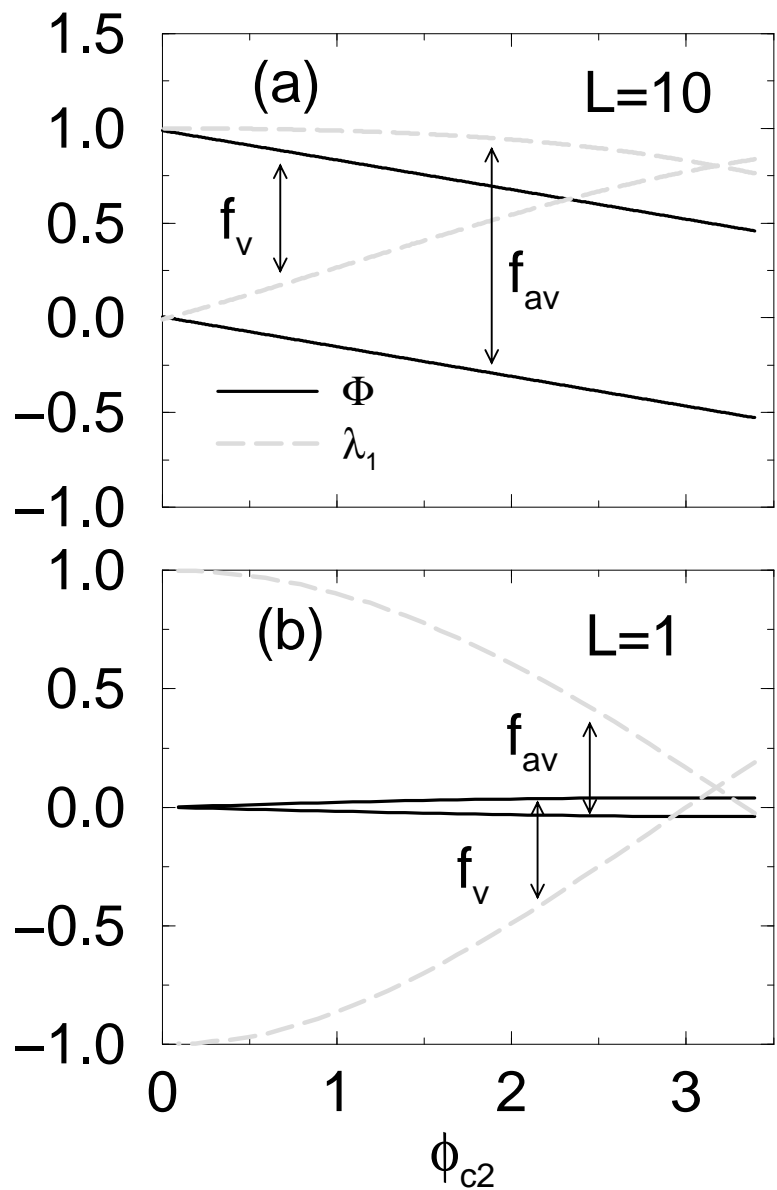
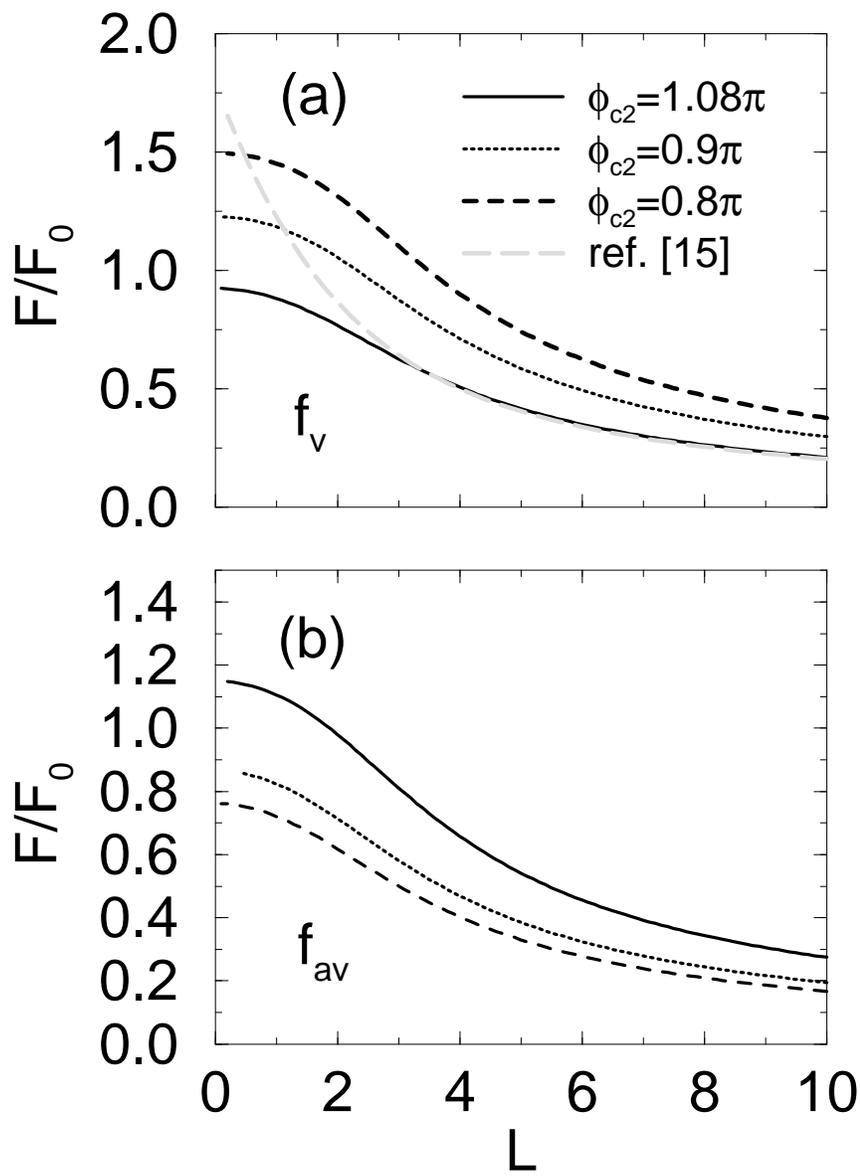
\end{document}